\documentclass[%
reprint,
superscriptaddress,
 amsmath,amssymb,
 aps,
prl,
]{revtex4-2}

\usepackage{graphicx}
\usepackage{dcolumn}
\usepackage{bm}
\usepackage[colorlinks=true,
            linkcolor=blue,
            citecolor=blue,
            urlcolor=blue
           ]{hyperref}
\usepackage{cleveref}

\usepackage{pdfpages} 
\usepackage{pgffor} 

\makeatletter
\AtBeginDocument{\let\LS@rot\@undefined}
\makeatother

\def\supplementfilename{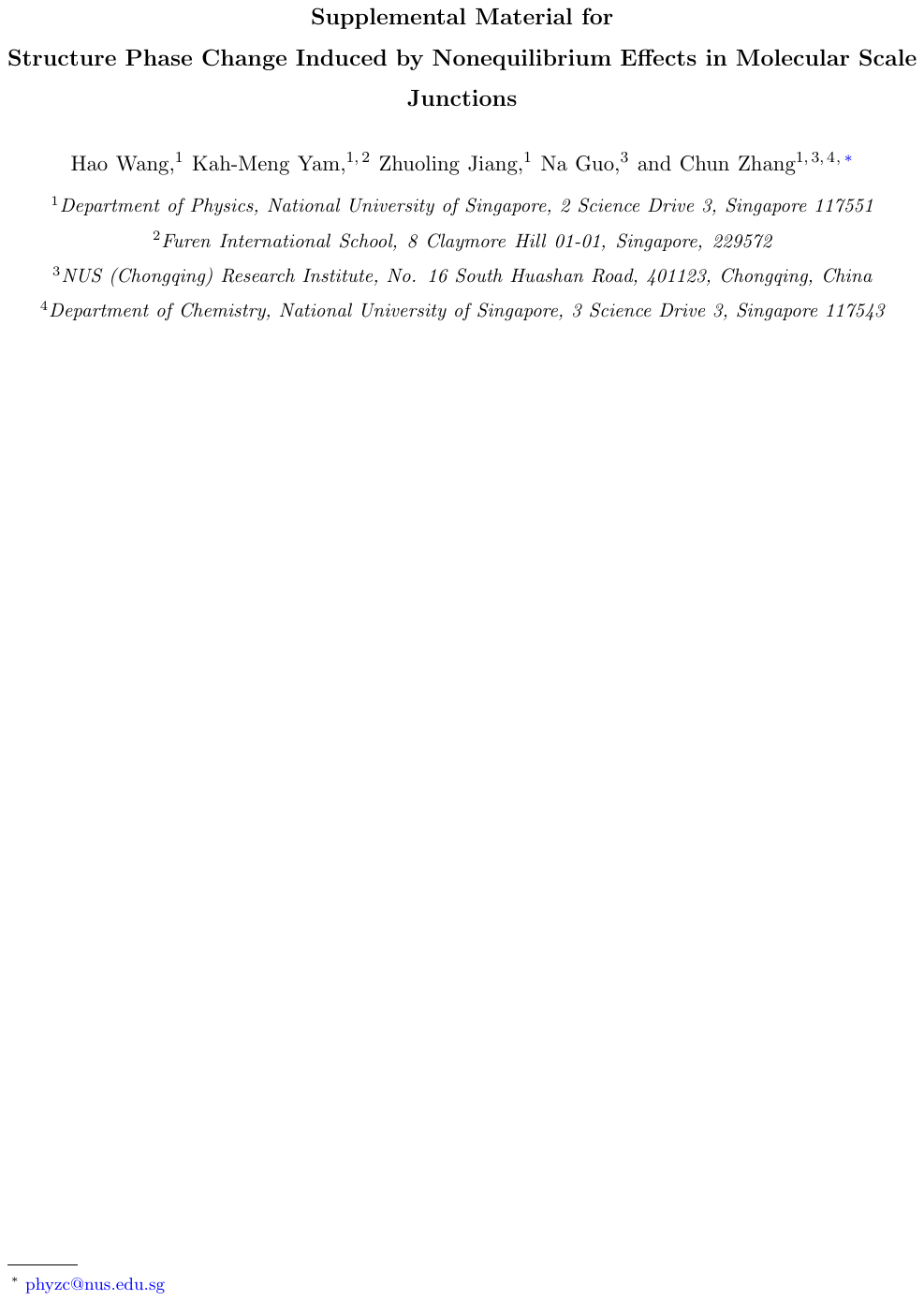}

\pdfximage{\supplementfilename}
\def\numbersupplementpages{\the\pdflastximagepages}

\newif\ifarXiv
\arXivtrue 

\begin{document}


\title{Structure Phase Change Induced by Nonequilibrium Effects in Molecular Scale Junctions}

\author{Hao Wang}
\affiliation{Department of Physics, National University of Singapore, 2 Science Drive 3, Singapore 117551}
\author{Kah-Meng Yam}
\affiliation{Department of Physics, National University of Singapore, 2 Science Drive 3, Singapore 117551}
\affiliation{Furen International School, 8 Claymore Hill 01-01, Singapore, 229572}
\author{Zhuoling Jiang}
\affiliation{Department of Physics, National University of Singapore, 2 Science Drive 3, Singapore 117551}
\author{Na Guo}
\affiliation{NUS (Chongqing) Research Institute, No. 16 South Huashan Road, 401123, Chongqing, China}
\author{Chun Zhang}
\email{phyzc@nus.edu.sg}
\affiliation{Department of Physics, National University of Singapore, 2 Science Drive 3, Singapore 117551}
\affiliation{NUS (Chongqing) Research Institute, No. 16 South Huashan Road, 401123, Chongqing, China}
\affiliation{Department of Chemistry, National University of Singapore, 3 Science Drive 3, Singapore 117543}


\begin{abstract}
The interrelationship between a material's structure and its properties lies at the heart of materials-related research. Finding how the changes of one affect the other is of primary importance in theoretical and computational materials studies. In this work, based on Hershfield nonequilibrium quantum statistics and the mean-field approach with steady-state density functional theory, we derive a first-principles method to calculate nonequilibrium effects induced forces acting on atoms, enabling structure optimizations and molecular dynamics simulations for molecular junctions under external biases. By applying the method to a few molecular devices, we found that in general, the external bias can induce profound nonequilibrium effects on both electronic/transport properties and the geometric structure of these devices, and consequent changes in electronic properties and geometric structure are closely interrelated. Particularly, when the bias voltage is above 1.0 V, significant structure phase changes could occur, causing dramatic changes in I-V characteristics and vibrational spectra. These findings greatly broaden our understanding of quantum electronic devices and provide a new avenue for discovering novel transport phenomena at molecular scale.
\end{abstract}

\maketitle

In the pursuit of miniaturized and high-performance electronic devices~\cite{HisME13,CPSME16,DESME19}, numerous molecular electronics have been proposed in recent decades, such as single-molecule transistors~\cite{MTE04,MOG09}, switches~\cite{LDMS04,TCMS07,SMJPS16}, diodes~\cite{ARDi03,LMSGNR11,TRSMD13} and many others~\cite{SpinT04,STQW05,SMM09,MWJ03,MEs03,SMJMC06}, all of which showed the power of controlling and promoting device functions with tunable electronic structures of molecular centers. Molecular electronic devices always operate under an external bias voltage. The bias-induced nonequilibrium effects present a great challenge for theoretical modeling of these devices. Computational approaches based on ground-state density functional theory (GS-DFT), such as the combination of GS-DFT and nonequilibrium Green's function (DFT+NEGF) techniques~\cite{NEGF_GH01,TranSIESTA02,SMEAGOL05} and the so-called multi-space constrained energy minimization method (MS-DFT)~\cite{MSDFT20}, have proven to be effective in qualitatively explaining electronic structures and transport properties of many molecular electronics with various degrees of success. Recently, it was shown that the steady-state density functional theory (SS-DFT)~\cite{SSDFT15}, which considers nonequilibrium effects in a dual-mean-field approach, is reliable in describing electronic and transport properties of molecular devices even when they are driven far away from equilibrium and thus beyond the GS-DFT based approaches~\cite{NeqE21}. 

Despite the success of the aforementioned methods in describing bias-induced nonequilibrium effects on electronic and transport properties of molecular devices, two important problems still remain unsolved: (1) how bias-induced nonequilibrium effects affect geometric structures of molecular devices and (2) how electronic properties and geometric structure of a molecular device correlate with each other under nonequilibrium conditions. 
When modeling molecular devices, a common practice is to assume that the structure of the device under a finite bias is the same as the zero-bias one that can be obtained by GS-DFT calculations, regardless of how significant the bias-induced changes in electronic and/or transport properties are. This seemingly unreasonable pre-assumption has been widely used in past two decades and its validity has never been seriously questioned due to the missing of a reliable theoretical method for calculating the forces acting on atoms under nonequilibrium conditions. 

In this work, we derive a first-principles method to calculate nonequilibrium effects induced forces acting on atoms by employing Hershfield nonequilibrium quantum statistics~\cite{SSNQS93} under the framework of SS-DFT. The method is then implemented into the SS-DFT package for structure optimizations and molecular dynamics simulations under finite biases. After applying the method to several molecular devices including an intriguing one recently fabricated in an experiment~\cite{SMTI22}, we show that the bias-induced nonequilibrium state has profound effects on both electronic/transport properties and geometric structures of molecular devices, and changes in geometric structure and electronic properties are closely related to each other. In particular, when the bias voltage is above 1 V, nonequilibrium effects induced forces may result in a significant structure phase change, which in turn causes dramatic changes in I-V characteristics and vibrational spectra of the devices. These findings clearly indicate the importance of bias-induced nonequilibrium effects on geometric structures of molecular devices that has been essentially neglected in almost all previous studies, and significantly improve our fundamental understanding of electron transport at molecular or nano scale.

\begin{figure}[t]
       \centering
       \includegraphics[width=0.9\linewidth]{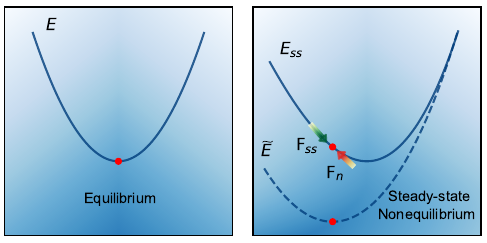}
       \vspace{-1em}
       \caption{Energy and force scheme in equilibrium and steady-state nonequilibrium. The bright red dots represent the structure of the device in equilibrium (left panel) and a steady-state nonequilibrium under a finite bias (right panel). }
       \label{fig:F1EF}
       \vspace{-1.5em}
\end{figure}

\textit{Nonequilibrium effects induced forces.}--In the conventional GS-DFT approach, the structure of a system can be obtained by optimizing the energy as shown in the left panel of Fig.~\ref{fig:F1EF}, which ends up with the ground-state structure (or the equilibrium structure at zero temperature). The GS-DFT based methods have proven to be efficient and accurate in determining the equilibrium structures of materials at various scales. In computational molecular electronics, the same GS-DFT based approach has been used to obtain the structure of the molecular device under zero bias, and then the structure is assumed to remain the same under an external bias voltage. This assumption neglects nonequilibrium effects induced forces acting on the molecular device and consequent structural changes. 

Bias-induced nonequilibrium effects on electronic properties are fully considered in the SS-DFT approach~\cite{SSDFT15} by employing the Hershfield nonequilibrium quantum statistics~\cite{SSNQS93}. It has been shown that the Hershfield nonequilibrium steady state of the device under a finite bias $V_b$ can be obtained by minimizing the effective energy $\widetilde{E}$, 
\begin{equation}\label{eq:Et}
       \widetilde{E}[\rho_t, \rho_n] = E_{\textsc{\it SS}}[\rho_t, \rho_n]-\frac{1}{2}eV_b\int d\boldsymbol{r}\rho_n(\boldsymbol{r}),
\end{equation}
where $E_{\textsc{\it SS}}$ is the regular energy of the transport steady state, which is a functional of two electron densities, the total electron density $\rho_t$ and the current-carrying (or nonequilibrium) electron density $\rho_n$. The following bias-dependent term, which is negative and measures how far the system is away from the equilibrium, is denoted as the nonequilibrium energy $E_n$. It has been proven that when the bias voltage $V_b$ goes to zero, the steady-state energy density functional $E_{\textsc{\it SS}}[\rho_t,\rho_n]$ reduces to the energy density functional $E[\rho_t]$ in GS-DFT, and consequently, SS-DFT reduces to conventional GS-DFT~\cite{SSDFT15}. 

In this work, we consider bias-induced nonequilibrium effects on geometric structure of the device. For this purpose, we introduce two forces acting on atoms in the device based on the above-mentioned energies (as shown in Eq.~\eqref{eq:NF}), the regular force $\mathbf{F_{\textsc{\it SS}}^I}$ acting on the atom $\mathbf{I}$, which is defined to be the first-order derivative of $E_{\textsc{\it SS}}$ with respect to the coordinates $\mathbf{R^I}$, and the nonequilibrium force $\mathbf{F_{\it n}^I}$, which originates from the derivative of $E_n$. The net force on atom $I$ is the summation of these two forces  $\mathbf{F_{\it net}^I}=\mathbf{F_{\textsc{\it SS}}^I}+\mathbf{F_{\it n}^I}$.
\begin{equation}
\begin{aligned}\label{eq:NF}
&\mathbf{F_{\textsc{\it SS}}^I}=-\frac{\partial E_{\textsc{\it SS}}\left[\rho_t, \rho_n\right]}{\partial \mathbf{R^I}}, \\
&\mathbf{F_{\it n}^I}=-\frac{\partial E_n}{\partial  \mathbf{R^I}}=\frac{eV_b}{2} \frac{\partial}{\partial \mathbf{R^I}}\left(\int d \boldsymbol{r} \rho_n(\boldsymbol{r})\right).
\end{aligned}
\end{equation}
Note that nonequilibrium effects are included in both of these two forces. Without an external bias, electrons are in equilibrium and net forces acting on atoms are zero. When a finite bias is applied, the nonequilibrium effects change the electronic structure, leading to nonzero regular forces $\mathbf{F_{\textsc{\it SS}}}$, and meanwhile nonequilibrium force $\mathbf{F_{\it n}}$ arises. These two forces together drive atoms away from the zero-bias equilibrium, and the system will finally arrive at a nonequilibrium steady-state structure in which $\mathbf{F_{\it n}}$ balances with $\mathbf{F_{\textsc{\it SS}}}$ (so the net force $\mathbf{F_{\it net}^I}$ becomes zero), as shown in the right panel of Fig.~\ref{fig:F1EF}. At this nonequilibrium steady state, the regular energy $E_{\textsc{\it SS}}$ is not at the minimum, while the effective energy $\widetilde{E}$ is. Numerical calculations of the two forces in Eq.~\eqref{eq:NF} are nontrivial. After some tedious but straightforward derivations, the two forces can be computed using density matrices calculated in SS-DFT. Force calculations, subsequent structure optimizations and nonequilibrium molecular dynamics (N-MD) simulations have been implemented into the SS-DFT package. Detailed computational methods and procedures can be found in Computational Methods section in Supplemental Material~\cite{SM} (see also references~\cite{SIESTA02,TM91,PBE96,DMF13,GTFDM14,MDNose84,MDHoover85,ASE17,PyNAO19,Pulay69} therein).

\begin{figure}[tbhp]
       \centering
       \includegraphics[width=0.95\linewidth]{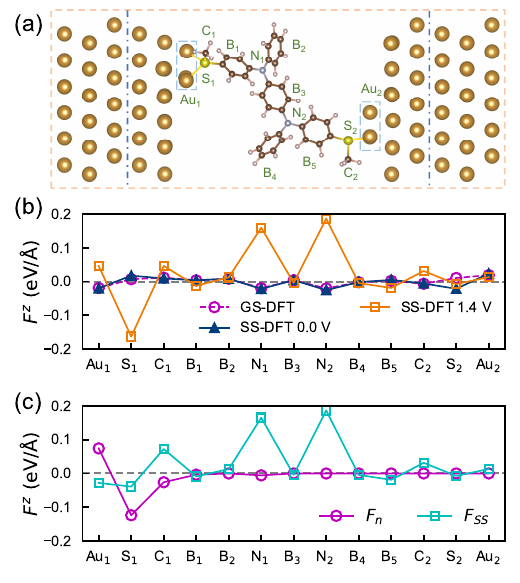}
       \vspace{-1em}
       \caption{(a) GS-DFT optimized geometric structure of the oligophenylene-bridged bis-triarylamines molecule device. Atoms in the device region 
       are divided into 13 groups, where Au$_1$ and Au$_2$ are Au atoms in left and right contacts that bond with the molecule, and B$_1$ to B$_5$ represent 5 phenyl rings from source to drain. (b) Average net forces on different atom groups along the transport direction ($z$-axis). 
       (c) Average nonequilibrium force ($\mathbf{F_{\it n}}$) and regular force ($\mathbf{F_{\textsc{\it SS}}}$) along the $z$-axis on different atom groups.}
       \label{fig:F2Fzn}
       \vspace{-1.5em}
\end{figure}

\textit{Oligophenylene-bridged bis-triarylamines molecule device.}--A recent experiment reported the fabrication and low-bias transport measurements of an intriguing molecular device that behaves like a single-molecule topological insulator~\cite{SMTI22}. The device consists of an oligophenylene-bridged bis-triarylamines molecule attached to gold electrodes through two sulfur atoms in contacts. 
Here we use the first-principles method discussed above to study the bias-induced nonequilibrium effects on the device's geometric structure and the interrelationship between the structure and electronic/transport properties. For this purpose, we build a device model as shown in \crefformat{figure}{Fig.~#2#1{(a)}#3}\cref{fig:F2Fzn}, which is initially optimized by GS-DFT calculations with two electrodes fixed to bulk structures. The SS-DFT calculation for this model yields the zero-bias conductance of 2.6$\times$ 10$^{-4}$ $\rm G_0$, 
which is in the same order as experimentally measured low-bias conductance (9.0$\times$ 10$^{-4}$ $\rm G_0$)~\cite{SMTI22}.  

\vspace{-1em}
We now apply our method to calculate forces acting on atoms under two bias voltages, 0 V and 1.4 V, and compare them with the conventional GS-DFT results. Hereinafter the forces calculated based on Eq.~\eqref{eq:NF} are referred to as SS-DFT forces. According to the widely used assumption that the structure does not change with the bias, the structure shown in \crefformat{figure}{Fig.~#2#1{(a)}#3}\cref{fig:F2Fzn} is used for both bias voltages. In \crefformat{figure}{Fig.~#2#1{(b)}#3}\cref{fig:F2Fzn}, at zero bias, SS-DFT net forces are essentially the same as the GS-DFT ones, and magnitudes of all calculated forces are within the force convergence criteria (0.02 eV/\AA), which is not surprising because the structure under study is obtained by GS-DFT structure optimization and the force calculations of SS-DFT at zero bias are essentially the same as GS-DFT. 
In contrast, when $V_b$ = 1.4 V, SS-DFT net forces show significant deviations from the GS-DFT ones. Especially, the forces on source contact Au and S atoms and two N atoms in middle that link different phenyl groups are much greater than the force convergence criteria. Similar trends are found for forces along the $x$ and $y$ directions as shown in Fig. S1. These differences are the consequence of the fact that under the bias of 1.4 V, the changes in electronic structures of these atoms are a lot more significant than other atoms, which can be seen from the differential charge density plot in Fig. S2. The calculated averaged regular forces $\mathbf{F_{\textsc{\it SS}}}$ and nonequilibrium forces $\mathbf{F_{\it n}}$ on each atom group under 1.4 V are illustrated in  \crefformat{figure}{Fig.~#2#1{(c)}#3}\cref{fig:F2Fzn}. The non-zero net forces on source contact Au and S atoms are mainly contributed by $\mathbf{F_{\it n}}$, while the origin of non-zero net forces on two middle N atoms is $\mathbf{F_{\textsc{\it SS}}}$. These results clearly suggest that at 1.4 V, these atoms shall not stay at their zero-bias equilibrium positions and consequently, the structure at 1.4 V shall be different from the structure at 0 V, so the widely used assumption that the bias voltage does not affect the geometric structure, in principle, is not correct. 

\begin{figure*}[th]
       \centering
       \includegraphics[width=0.9\linewidth]{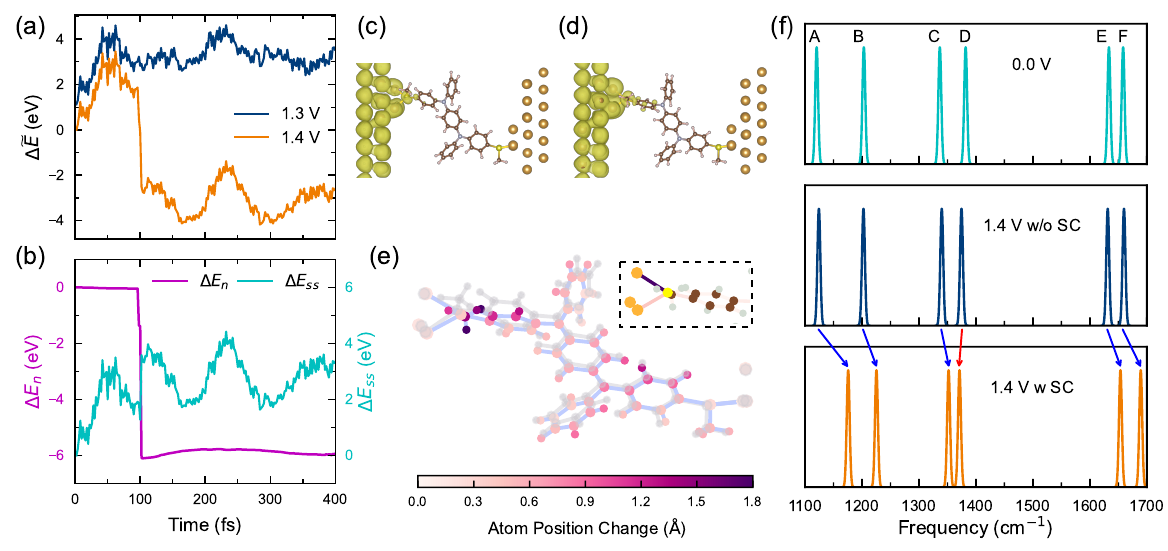}
       \vspace{-1em}
       \caption{Electronic and structural properties. (a) Variations of effective energy at 1.3 V and 1.4 V. $\widetilde{E}$ at 0 fs of 1.4 V is set to 0 eV. (b) Variations of nonequilibrium energy (left axis in magenta) and regular energy (right axis in cyan) at 1.4 V. (c),(d) Iso-surface plot of nonequilibrium charge density for the initial structure and the structure right after 100 fs. (e) Changes in atom position and bond length. The gray structure is the initial GS-DFT optimized structure and the pink one is the structure by an annealing process after 300 fs of N-MD. The inset shows bond length variations in the source contact region, where the color is scaled by a factor of 6. (f) Vibrational spectra for six selected Raman modes of the molecule without structure changes (w/o SC) at 0 V (top) and 1.4 V (middle) and with structure changes (w SC) at 1.4 V (bottom).}
       \label{fig:F3SC}
       \vspace{-1.5em}
\end{figure*}

\textit{Nonequilibrium effects induced structure phase change.} --With the calculated forces, we are able to optimize the structure of the device under a finite bias. Due to the complicated effective energy surface of the given system, 
we determine the structure by ab initio nonequilibrium molecular dynamics (N-MD) simulations under room temperature (300 K) together with a follow-up annealing process. For all N-MD simulations under different bias voltages, the starting structure is chosen to be the one optimized at zero bias. As shown in Fig. S3, at various bias voltages less than 1.4 V, effective energy variations with N-MD simulation time are all essentially the same as the energy variations obtained from the GS-DFT based MD simulations at zero bias, indicating that no significant bias-induced structure changes occur during these simulations. Something interesting happens at 1.4 V. In \crefformat{figure}{Fig.~#2#1{(a)}#3}\cref{fig:F3SC}, before 100 fs, the evolutions of effective energy for 1.3 V and 1.4 V are qualitatively the same, while after 100 fs, the effective energy for the case of 1.4 V dramatically drops by around 4.5 eV, clearly demonstrating that a structure phase change occurs. As shown 
in \crefformat{figure}{Fig.~#2#1{(b)}#3}\cref{fig:F3SC}, when the structure changes after 100 fs, $E_{\textsc{\it SS}}$ increases and $E_n$ decreases, so the drop of the total effective energy and in turn the structure phase change are mainly caused by the variation of $E_n$. 

The increase of $E_{\textsc{\it SS}}$ is a natural consequence of the fact that the starting structure is the zero-bias one obtained by GS-DFT. When the structure of the system is driven away from the ground-state structure, certainly its $E_{\textsc{\it SS}}$ increases. The increase of $E_{\textsc{\it SS}}$ is about 1.5 eV, suggesting that the structure phase change is quite significant. To understand the drop of $E_n$, which depends on the total number of nonequilibrium electrons, we plot iso-surfaces of nonequilibrium electron density for the initial structure and the one right after 100 fs in \crefformat{figure}{Fig.~#2#1{(c)}#3}\cref{fig:F3SC} and \crefformat{figure}{#2#1{(d)}#3}\cref{fig:F3SC}. After the structure phase change, the number of nonequilibrium electrons at the source contact significantly increases, leading to the dramatic drop of $E_n$ and in turn the decrease of the total effective energy despite the increase of $E_{\textsc{\it SS}}$. \crefformat{figure}{Fig.~#2#1{(e)}#3}\cref{fig:F3SC} shows the comparison between the initial structure and the one with the structure phase change, which is obtained by an annealing process after 300 fs of N-MD simulation. The structure changes mainly occur at the source contact, where the average length of Au-S bonds is elongated by about 0.3 \AA, and the first methyl and phenyl group (C$_1$ and B$_1$ in \crefformat{figure}{Fig.~#2#1{(a)}#3}\cref{fig:F2Fzn}) are rotated about the $z$-axis approximately 30$^{\circ}$ and 25$^{\circ}$, respectively. An animation for the structure phase change in N-MD simulations that occurs around 100 fs can be found in Supplemental Material, where the structure change at the source contact can be clearly seen. 

The structure phase change leads to changes in vibrational modes of this molecular device, which possibly can be detected in experiments by surface-enhanced Raman spectroscopy measurements~\cite{VibSMJ14,BIEDS21,BSMJ19}. The Raman spectrum of the central molecule is illustrated in Fig. S6 and six of the most active vibrational modes are selected to assess the bias-induced nonequilibrium effects on these modes. From the vibrational frequencies in \crefformat{figure}{Fig.~#2#1{(f)}#3}\cref{fig:F3SC} minor shifts in frequency are observed at 1.4 V without structure change (w/o SC) as compared to the zero-bias ones. However, after accounting for the structure phase change, five vibrational modes show significant blueshifts, while mode D undergoes a redshift of about 11 cm$^{-1}$. As the structure changes occur primarily at the source contact, mode A, which involves the vibration of the source contact S atom and the connected phenyl group, exhibits the largest blueshift up to 55 cm$^{-1}$. These analyses based on Raman spectra offer a valuable approach to experimentally verify the predicted bias-driven nonequilibrium structure phase change of the device.

\begin{figure}[t]
       \centering
       \includegraphics[width=\linewidth]{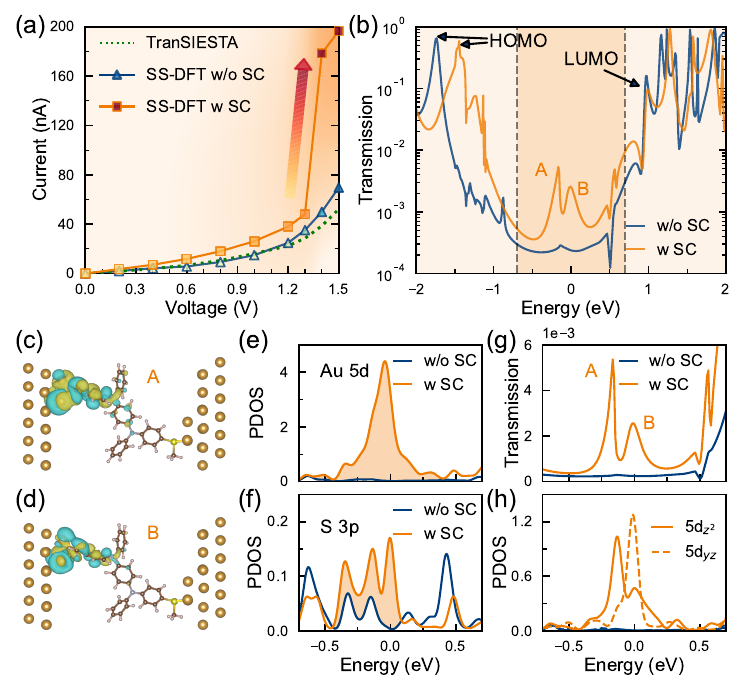}
       \vspace{-2em}
       \caption{Transport properties of the molecular junction. (a) I-V characteristics of TranSIESTA (green), SS-DFT without structure changes (w/o SC, blue) and SS-DFT with structure changes (w SC, orange). (b) Transmission spectra of SS-DFT w/o and w SC at 1.4 V. The bias window is denoted by two vertical dotted lines. (c), (d) Tunnelling eigenchannels for peak A and B in (b). (e), (f) PDOS at 1.4 V w/o and w SC of 5d orbitals of three Au atoms at source contact binding with the S atom and 3p orbitals of the S atom. The highlighted differences in orange signify the enhancement of coupling between Au and S atoms. (g) Transmission spectra in the bias window. (h) PDOS of 5d$_{z^2}$ (orange solid line) and 5d$_{yz}$ (orange dashed line) of source contact Au atoms.}
       \label{fig:F4Tran}
       \vspace{-2em}
\end{figure}

\textit{Effects of structure phase change on transport properties.}--To elucidate effects of structure phase change on transport properties of the device, we calculate I-V curve of the device by performing GS-DFT based method (TranSIESTA) and SS-DFT method with the assumption that the structure does not change with the bias (w/o SC), and SS-DFT method with structure optimizations done by N-MD simulations (w SC). The results are shown in \crefformat{figure}{Fig.~#2#1{(a)}#3}\cref{fig:F4Tran}, where the I-V curve obtained from SS-DFT w/o SC is essentially the same as that from TranSIESTA at low biases ($V_b\ < $ 1.0 V), while when $V_b\ >$ 1.0 V, the SS-DFT generates slightly higher currents than TranSIESTA. In contrast, when considering the structure change due to the nonequilibrium effects induced forces, the current from SS-DFT starts to deviate from the TranSIESTA one at low biases around 0.4 V, and strikingly, at 1.4 V when the structure phase change occurs, the current surges up from around 40 nA to about 180 nA, clearly indicating strong effects of the structure phase change on the transport properties. 

To shed light on the origin of this current surge, we plot in \crefformat{figure}{Fig.~#2#1{(b)}#3}\cref{fig:F4Tran} the transmission spectra from SS-DFT calculations under 1.4 V w/o SC and w SC. 
The transmission peak through the LUMO (see Fig. S4) of the molecule essentially remains the same, while the HOMO transmission peak is shifted towards the bias window by approximately 0.3 eV after the structure phase change occurs, leading to higher transmissions in the bias window. Interestingly, for the case of w SC, two additional transmission peaks arise in the middle of the bias window, one at -0.17 eV (peak A) and the other one at -0.01 eV (peak B). These two transmission peaks clearly are the main reasons that the electric current surges up after the structure changes. Since the two peaks appear in the middle of the bias window that are far away from narrow HOMO and LUMO peaks, it is reasonable to assume that the electron transport at the energies of these two peaks can be approximated by electron tunnelling through a featureless potential barrier provided by the molecule. Therefore, the coupling between the molecule and two metal electrodes (instead of molecule states) would be the decisive factor for the transport.

The tunnelling eigenstate analysis for peak A and B (\crefformat{figure}{Fig.~#2#1{(c)}#3}\cref{fig:F4Tran} and \crefformat{figure}{#2#1{(d)}#3}\cref{fig:F4Tran}) suggests that the coupling at the source contact plays a dominant role. The calculated Projected Density of States (PDOS) for Au atoms (bonded with S) and the S atom (Fig. S5) at the source contact clearly show that the structure phase change significantly enhances the coupling between Au 5d and S 3p orbitals (\crefformat{figure}{Fig.~#2#1{(e)}#3}\cref{fig:F4Tran} and \crefformat{figure}{#2#1{(f)}#3}\cref{fig:F4Tran}) in the middle of bias window from around -0.5 to 0.1 eV. After projecting the tunnelling eigenchannels to different Au 5d orbitals (see Table S1), we see that the transmission peak A is dominated by tunnelling through Au 5d$_{z^2}$ and peak B is mainly governed by tunnelling through both 5d$_{yz}$ and 5d$_{z^2}$, which can be confirmed by comparing the transmission spectra in bias window (\crefformat{figure}{Fig.~#2#1{(g)}#3}\cref{fig:F4Tran}) and PDOS of source contact Au 5d$_{yz}$ and 5d$_{z^2}$ (\crefformat{figure}{Fig.~#2#1{(h)}#3}\cref{fig:F4Tran}). As a short summary, the structure phase change under 1.4 V significantly enhances the coupling between the Au atoms and the S atom at source contact, and then dramatically increases the electric current tunnelling through the molecule. 

Additional examinations are carried out on two widely studied molecular junctions, namely gold-1,4-benzenedithiolate and gold-oligophenylenediamine. For both cases, I-V curves from TranSIESTA (Fig. S7) are consistent with previous studies ~\cite{AuNMJ17,MJAgE13,MMSMJ16,GW11}. Under low biases, the currents obtained from SS-DFT w/o SC are essentially the same as those from TranSIESTA. However, after accounting for structure changes, electric currents increase significantly when the bias is greater than 1 V. These calculations clearly suggest that the predicted structure phase change and structure-change induced surge of current could be general for most molecular scale junctions. Such high current surge for sure generates huge amount of heat and causes the instability of the devices (see more discussions in Supplemental Material for effects of current surge on thermal properties and stability), which may shed light on the experimental observations that many molecular devices become unstable and even break down at bias voltage above 1 V~\cite{SSMD09,BDSMJ15,BDTBD08}.

\textit{Conclusion.}--In this work, we derive a first-principles method to calculate forces acting on atoms under steady-state nonequilibrium conditions. The method is then implemented into SS-DFT package for structure optimizations and nonequilibrium molecular dynamics simulations. By applying the method to several molecular devices, we show that when the external bias is above 1 V, nonequilibrium effects could induce a significant structure phase change in the molecular scale junction, which causes dramatic changes in I-V curve and vibrational spectra of the device. 
These findings greatly broaden our understanding of nonequilibrium electron transport at molecular scale and offer a new avenue for the future computational design of novel molecular devices. We believe that our method provides a valuable tool to investigate nonequilibrium phenomena in some other systems such as electrochemical cells and scanning tunnelling microscopes.

\begin{acknowledgments}
We acknowledge the support from Ministry of Education of Singapore (A-0004216-00-00). The calculations were performed on computational facilities of the National Supercomputing Centre (NSCC), Singapore.
 
\end{acknowledgments}

\bibliography{ssdft_neq}

\ifarXiv
    

\section*{Supplementary material}

# I. COMPUTATIONAL METHODS

## A. Nonequilibrium effects induced forces

Numerical computations of regular and nonequilibrium forces are nontrivial. In SS-DFT, the total electron density $\rho^t$ is calculated as the summation of equilibrium electron density $\rho^e$ and nonequilibrium electron density $\rho^n$, which are determined by two corresponding mean-field Hamiltonians $\mathcal{H}^e$ and $\mathcal{H}^n$ [1]. Using the NEGF techniques [2, 3], these two densities can be computed by the Green's functions as follows,

$$\rho^e = -\frac{1}{\pi}\text{Im}\left[\int_{-\infty}^{\mu_R} G^e\left(\varepsilon\right)d\varepsilon\right],\tag{1}$$

$$\rho^n = \frac{1}{\pi}\left[\int_{\mu_R}^{\mu_L} G^n\left(\varepsilon\right)\Gamma_L\left(\varepsilon\right)\left(G^n\left(\varepsilon\right)\right)^{\dagger}d\varepsilon\right],\tag{2}$$

where $\mu_L$ and $\mu_R$ are the chemical potentials of source and drain. The Green's functions can be calculated using the corresponding Hamiltonian as $G\left(\varepsilon\right) = (\varepsilon S - \mathcal{H} - \Sigma_L\left(\varepsilon\right) - \Sigma_R\left(\varepsilon\right))^{-1}$ where $S$ is the overlap matrix of basis functions and $\Sigma_{L/R}$ are self-energies of source/drain. The coupling function $\Gamma_L$ is defined to be $\frac{i}{2}\left(\Sigma_L\left(\varepsilon\right) - \Sigma_L^{\dagger}\left(\varepsilon\right)\right)$. When calculating the forces with the localized basis set, the Pulay force [4] (that originates from the variations of localized basis functions when atoms change positions) needs to be considered. Considering the two-density dependences of density functionals in SS-DFT and following the same spirit of GS-DFT based force calculations, after some tedious derivations, the two forces under nonequilibrium conditions in SS-DFT can be computed with density matrices as follows,

$$\begin{aligned}
\mathbf{F}_{SS}^{\mathbf{I}} = & \sum_{\mu\nu}\left(\Omega_{\nu\mu}^e\frac{\partial S_{\mu\nu}}{\partial\mathbf{R}^{\mathbf{I}}} + \Omega_{\nu\mu}^n\frac{\partial S_{\mu\nu}}{\partial\mathbf{R}^{\mathbf{I}}}\right) - \frac{\partial E_{NN}}{\partial\mathbf{R}^{\mathbf{I}}} \\
& - \sum_{\mu\nu}\frac{\partial T_{\mu\nu}}{\partial\mathbf{R}^{\mathbf{I}}}\rho_{\nu\mu}^t - \sum_{\mu\nu}\left\langle\phi_\mu\left|\frac{\partial V_{ext}}{\partial\mathbf{R}^{\mathbf{I}}}\right|\phi_\nu\right\rangle\rho_{\nu\mu}^t \\
& + 2\sum_{\mu\nu}\left\langle\phi_\mu\left|V_{ext} + V_H\right|\nabla\phi_\nu\right\rangle\rho_{\nu\mu}^t \\
& + 2\sum_{\mu\nu}\left\langle\phi_\mu\left|V_{xc}^e\right|\nabla\phi_\nu\right\rangle\rho_{\nu\mu}^e + 2\sum_{\mu\nu}\left\langle\phi_\mu\left|V_{xc}^n\right|\nabla\phi_\nu\right\rangle\rho_{\nu\mu}^n,
\end{aligned}\tag{3}$$

$$\mathbf{F}_n^{\mathbf{I}} = \frac{eV_b}{2}\sum_{\mu\nu}\frac{\partial S_{\mu\nu}}{\partial\mathbf{R}^{\mathbf{I}}}\rho_{\nu\mu}^n.\tag{4}$$

In above equations, $S_{\mu\nu} = \langle\phi_\mu|\phi_\nu\rangle$ is the overlap matrix element between basis functions $\phi_\mu$ and $\phi_\nu$, $V_H$ is the Hartree potential, $V_{xc}^e$ and $V_{xc}^n$ are exchange-correlation potentials for equilibrium and nonequilibrium electrons, respectively. The derivatives of the kinetic matrix $T_{\mu\nu}$, the external

potential $V_{ext}$ and nuclear-nuclear interaction energy $E_{NN}$ can be calculated in a similar way to GS-DFT calculations in SIESTA [5] while taking into account the two-density dependence of SS-DFT. Two energy density matrices $\Omega^e$ and $\Omega^n$ are defined as

$$\Omega^e = -\frac{1}{\pi} \text{Im} \left[ \int_{-\infty}^{\mu_R} \varepsilon G^e(\varepsilon) d\varepsilon \right], \tag{5}$$

$$\Omega^n = \frac{1}{\pi} \left[ \int_{\mu_R}^{\mu_L} \varepsilon G^n(\varepsilon) \Gamma_L(\varepsilon) (G^n(\varepsilon))^\dagger d\varepsilon \right]. \tag{6}$$

These equations have been implemented into SS-DFT package to calculate the nonequilibrium effects induced forces and to perform structure optimizations and nonequilibrium molecular dynamics simulations under an external bias voltage.

## B. Computational details

Structure optimizations and transport calculations under the framework of GS-DFT were done with SIESTA [5] and TranSIESTA [3], respectively. Nonequilibrium effects induced forces and subsequent structure optimizations and transport calculations were performed with SS-DFT [1]. In all calculations, norm-conserving pseudopotentials generated with Troullier-Martins scheme [6] were employed and scalar-relativistic effects were considered for Au. The generalized gradient approximation (GGA) of exchange-correlation functional in Perdew-Burke-Enzerhof (PBE) [7] format was used for GS-DFT calculations, and the GGA in PBE with nonequilibrium corrections [8, 9] was adopted in SS-DFT. The double-$\zeta$ polarized (DZP) basis sets were used for all elements. The convergence criteria for the density matrix, energy and force in the calculations were set to be $10^{-4}$, $10^{-4}$ eV and 0.02 eV/Å, respectively. To conserve the charge in the device region, the tolerance of charge correction was set to be 0.005 electrons. The constrained nonequilibrium molecular dynamics (N-MD) simulations were carried out in the canonical ensemble using the Nosé-Hoover thermostat approach [10, 11] with 2 fs time step at 300 K. In all N-MD simulations, only the linker Au atoms and the central molecule were permitted to move. The Raman spectrum was calculated using ASE [12] and PyNAO package [13]. The force constant matrix was calculated using the finite displacement method based on the forces in Eq. (2) and then vibrational frequencies and modes were obtained by SIESTA's Vibra package.

## II.  THE STABILITY OF THE MOLECULAR DEVICE AFTER THE STRUCTURE PHASE CHANGE

The predicted surge of the current at 1.4 V could severely affect the stability of the device due to the heat generated by the much higher electric current. The heat capacity ($C_V$) of the system can be estimated by the variance of effective energy in N-MD simulations as

$$C_V = \frac{<\widetilde{E}^2> - <\widetilde{E}>^2}{k_B T^2},$$ (7)

where $k_B$ is the Boltzmann constant and $T$ is the temperature, which is equal to 300 K in our calculations. As shown in Fig. 3(a), the effective energy fluctuation after the energy drop is significantly higher at 1.4 V than that at 1.3 V, indicating a significant change of $C_V$ after the structure change. Using the effective energies from 150 fs to 300 fs, the calculated $C_V$ from GS-DFT and SS-DFT at 1.3 V are in close agreement, with values of 0.031 and 0.032 eV/K, respectively. However, under the bias voltage of 1.4 V at which the structure phase change occurs, the $C_V$ rises to 0.084 eV/K. Assuming that the system can maintain its temperature at 1.3 V, the additional heat generated after some time $t$ (compared with that of 1.3 V) by the increased electric current at 1.4 V can be computed as $Q = (I - I_0) V t$, where $I_0$ is the electric current at 1.3 V. If the heat dissipation of the device at 1.4 V is the same as at 1.3 V, the change of temperature can be roughly estimated by the following equation,

$$Q = (I - I_0) V t = C_V \Delta T,$$ (8)

where $\Delta T$ is the change of temperature. Our calculation yields a high temperature rising rate of 13.6 K/ps, suggesting that under 1.4 V, the temperature of the system could rise to a high value at a very short period of time, causing the device unstable. The greatly weakened stability of the molecular device under 1.4 V could result in intense molecular vibrations, leading to drastic oscillations of electric current or even the breakdown of the device.

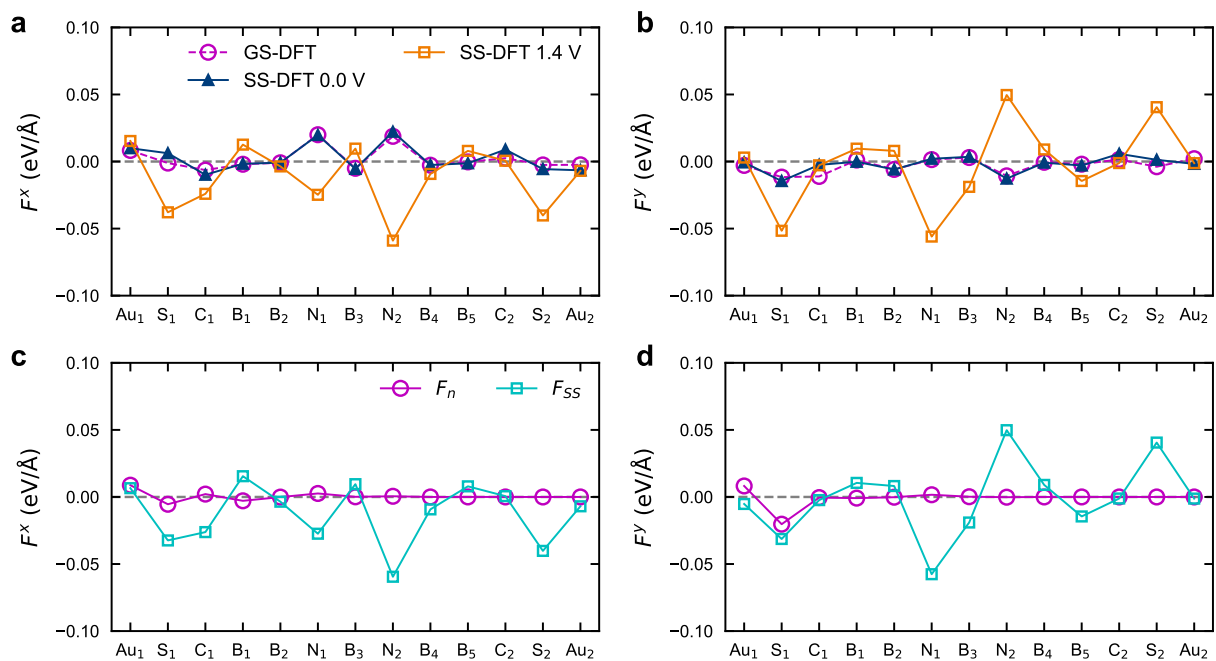

FIG. S1. Average net force (upper panel) and nonequilibrium force ($\mathbf{F}_n$), regular force ($\mathbf{F}_{SS}$) (lower panel) on group atoms along $x$-axis (a, c) and $y$-axis (b, d).

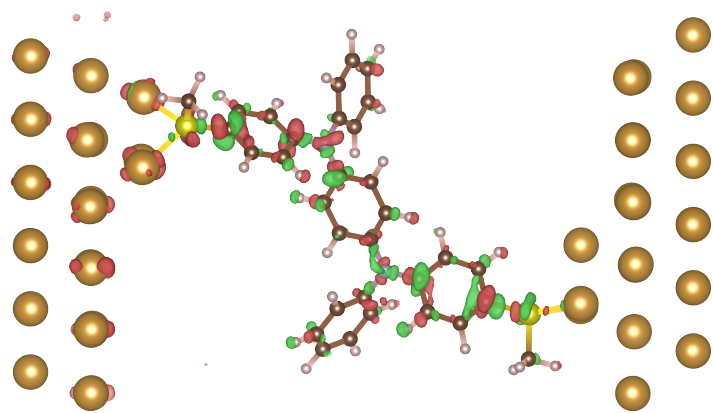

FIG. S2. Bias-induced total charge density change, $\Delta\rho_t = \rho_t(1.4V) - \rho_t(0.0V)$. The red (green) isosurfaces represent charge accumulation (depletion).

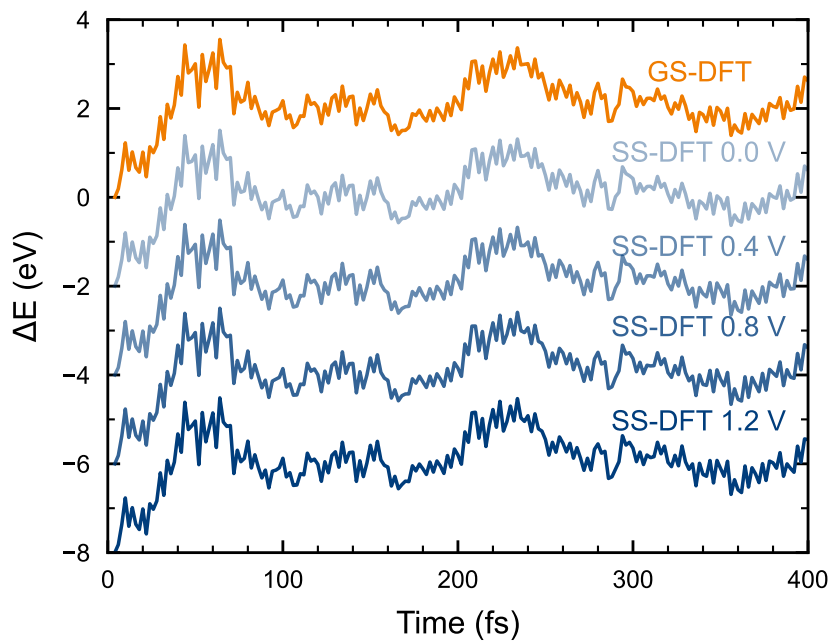

FIG. S3. Variations in total (effective) energies at 300 K from GS-DFT (orange) and SS-DFT (blue) based N-MD simulations. The total (effective) energies at 0 fs are set to be the zero points and energies at 0.0, 0.4, 0.8 and 1.2 V are shifted down by 2, 4, 6 and 8 eV, respectively.

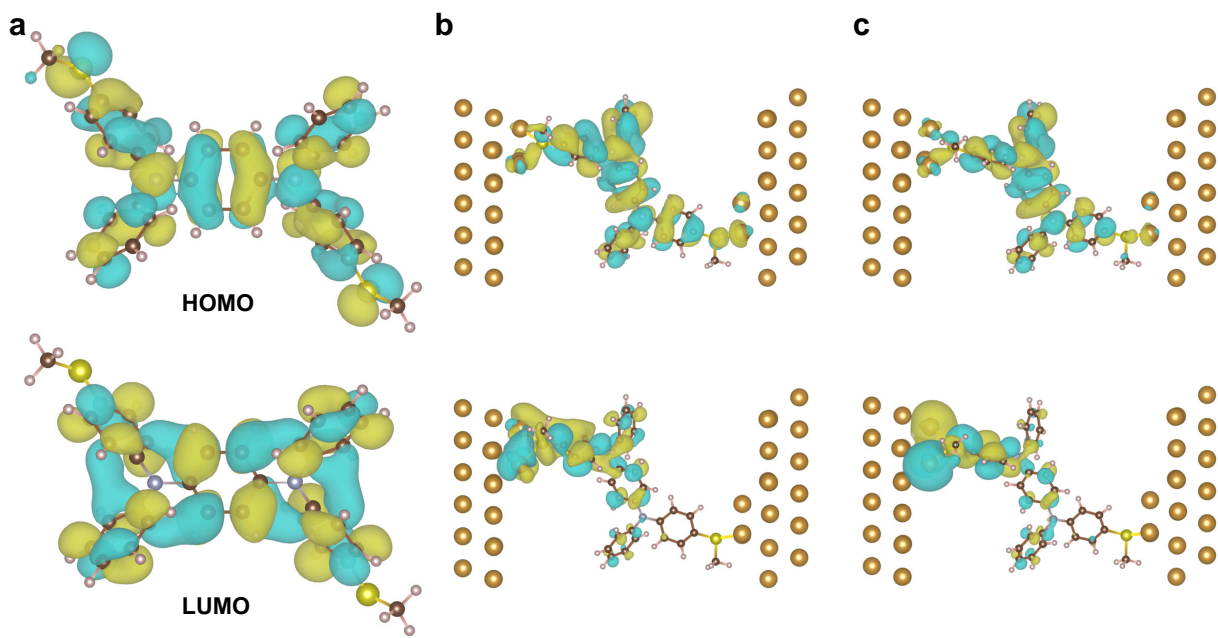

FIG. S4. Molecule orbitals of the isolated molecule (a) and tunnelling eigenchannels of corresponding HOMO (upper panel) and LUMO (lower panel) peaks (as indicated in Fig. 5b) without (b) and with (c) structure changes at 1.4 V.

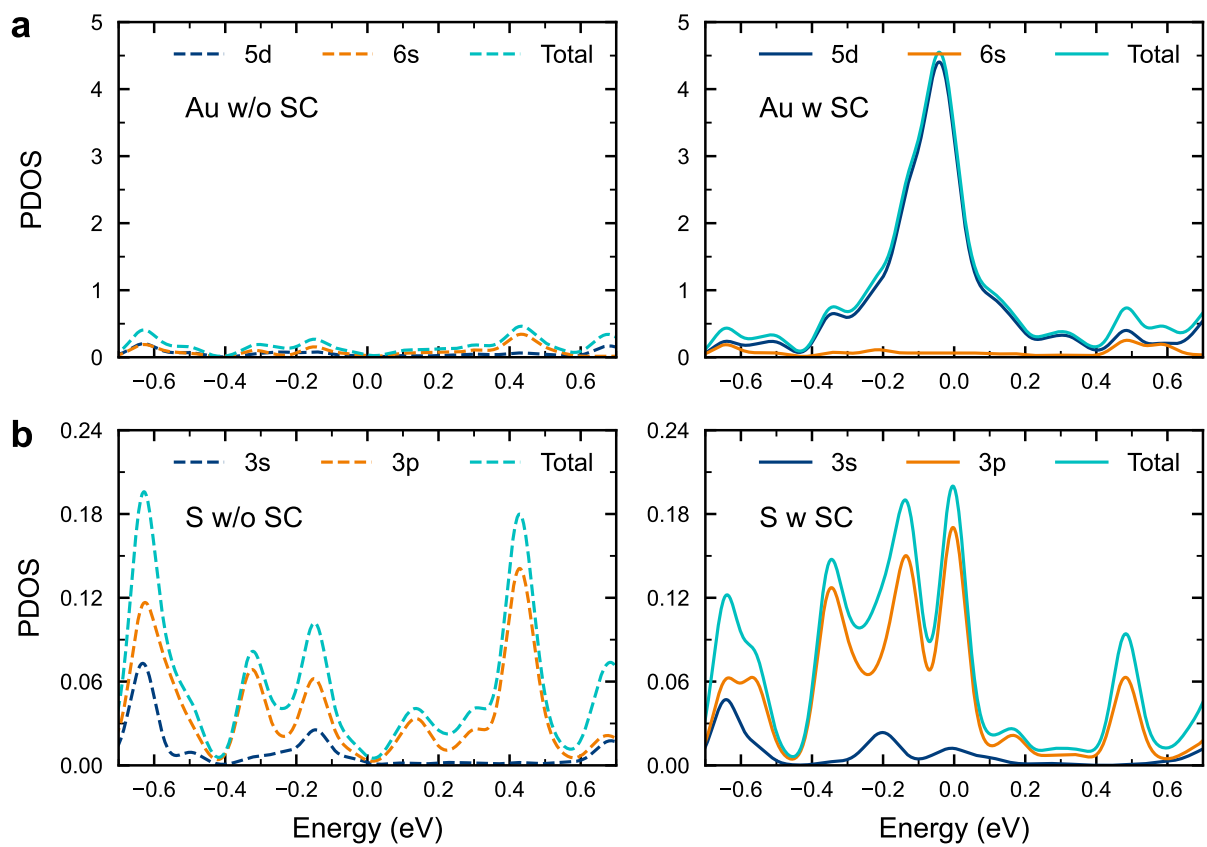

FIG. S5. Projected density of states at 1.4 V without (left dashed line) and with (right solid line) structure changes (SC) of Au atoms (a) and the S atom (b).

TABLE SI. Normalized occupation of five 5d orbitals of Au atoms for tunnelling eigenchannels A and B.

| Occupation | $5d_{xy}$ | $5d_{yz}$ | $5d_{z^2}$ | $5d_{xz}$ | $5d_{x^2-y^2}$ |
|---|---|---|---|---|---|
| A | 0.18 | 0.08 | 0.38 | 0.12 | 0.25 |
| B | 0.17 | 0.24 | 0.26 | 0.18 | 0.14 |

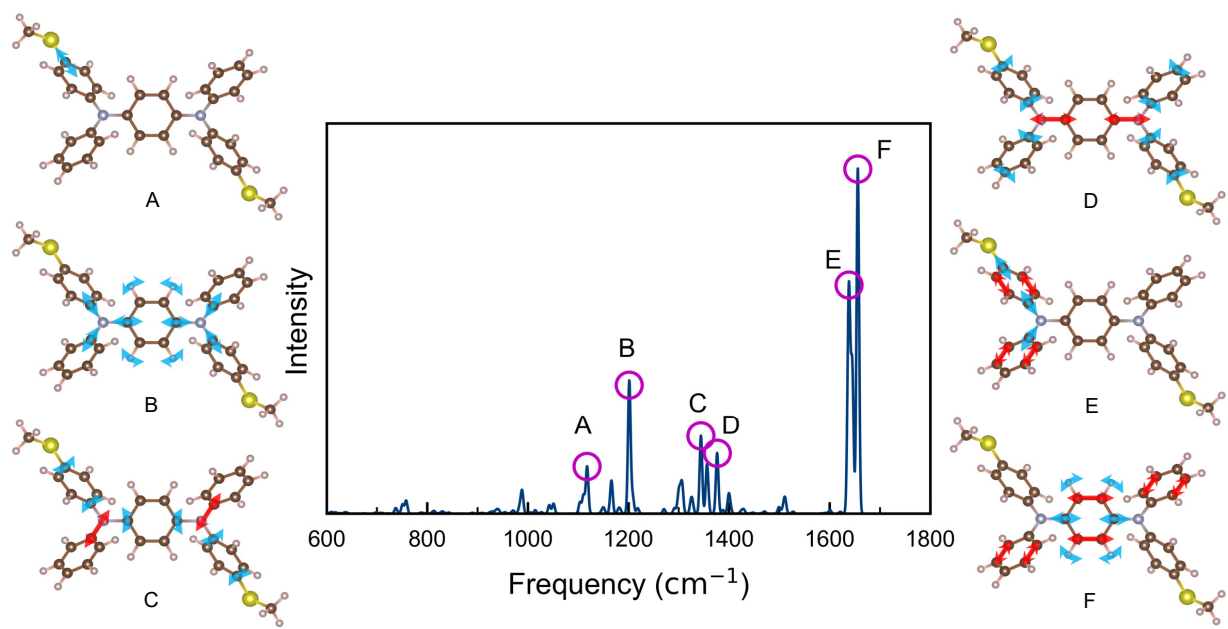

FIG. S6. Raman spectrum of the isolated molecule with Gaussian broadening. Six selected frequencies are highlighted by magenta circles and the vibrational modes are shown around.

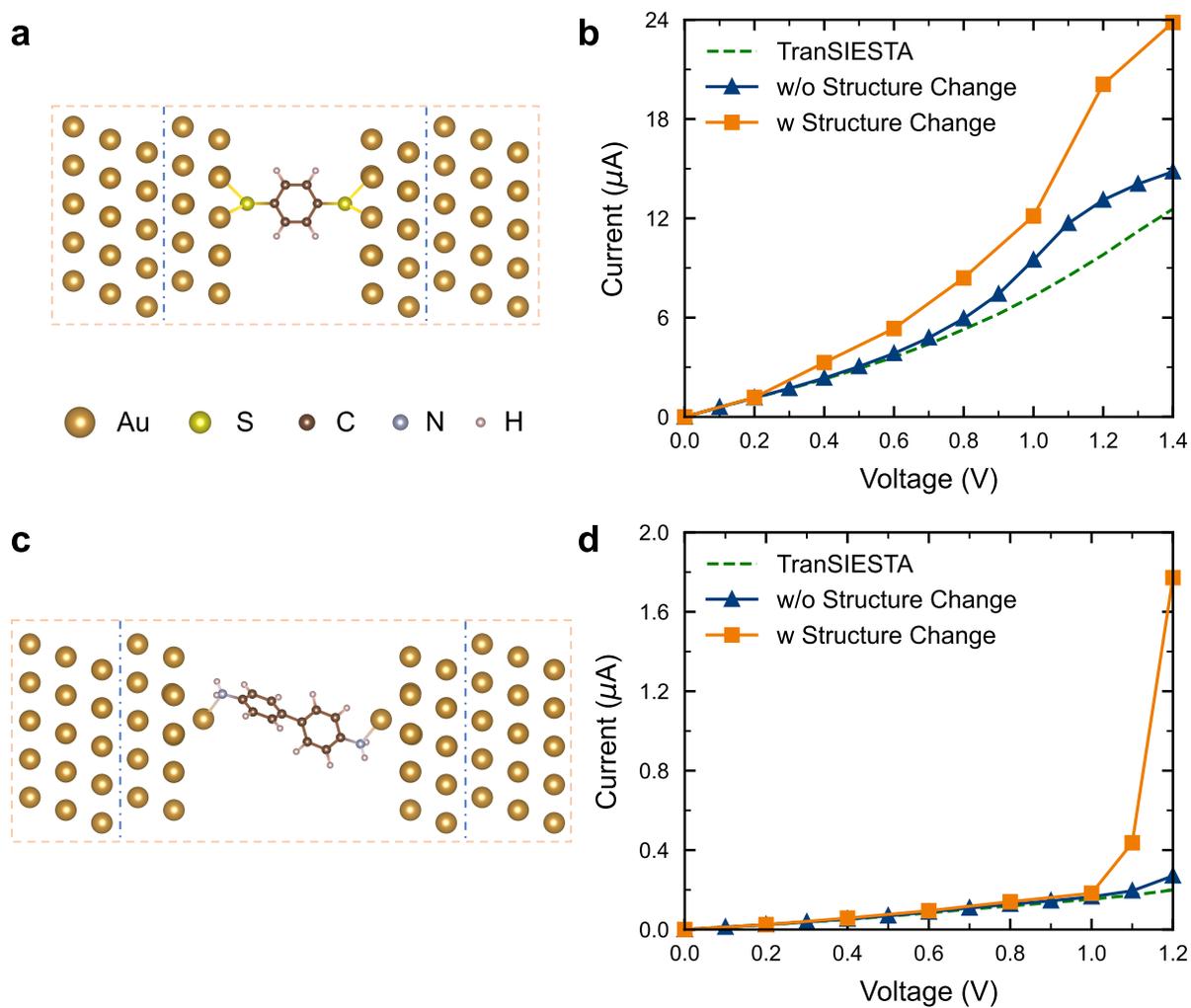

FIG. S7. Transport properties of the benzenedithiolate and oligophenylenediamine molecular junction. (a), (b) Atomic structures of gold-1,4-benzenedithiolate and gold-oligophenylenediamine molecular junction. (c), (d) I-V characteristics of TranSIESTA (green dotted line), SS-DFT without (w/o) structure changes (blue line with triangle marker) and SS-DFT with (w) structure changes (orange line with square marker) of the two devices.


\fi

\end{document}